\documentclass[journal=jacsat,manuscript=article]{achemso}


\usepackage[version=3]{mhchem} % Formula subscripts using \ce{}
\usepackage{siunitx}
\usepackage{color,soul}
\usepackage[colorinlistoftodos, color=green!20, textsize=scriptsize, prependcaption]{todonotes}
\usepackage{graphicx} 
\usepackage{epstopdf}
\usepackage{url}
\author{Young Won Woo}
\affiliation{Department of Materials, Imperial College London, London SW7 2AZ, UK}
\alsoaffiliation{Department of Materials Science and Engineering, Yonsei University, Seoul 03722, Korea}

\author{Zhenzhu Li}
\affiliation{Department of Materials, Imperial College London, London SW7 2AZ, UK}

\author{Young-Kwang Jung}
\affiliation{Department of Chemical Engineering \& Biotechnology, University of Cambridge, Cambridge CB3 0AS, UK}

\author{Ji-Sang Park}
\affiliation{SKKU Advanced Institute of Nanotechnology and Department of Nano Engineering, Sungkyunkwan University, Suwon 16419, Korea}

\author{Aron Walsh}
\email{a.walsh@imperial.ac.uk}
\affiliation{Department of Materials, Imperial College London, London SW7 2AZ, UK}
\title[]
  {Inhomogeneous Defect Distribution in Mixed-Polytype Metal Halide Perovskites}


\begin{document}
\setlength{\marginparwidth}{2cm}

\begin{tocentry}
\includegraphics{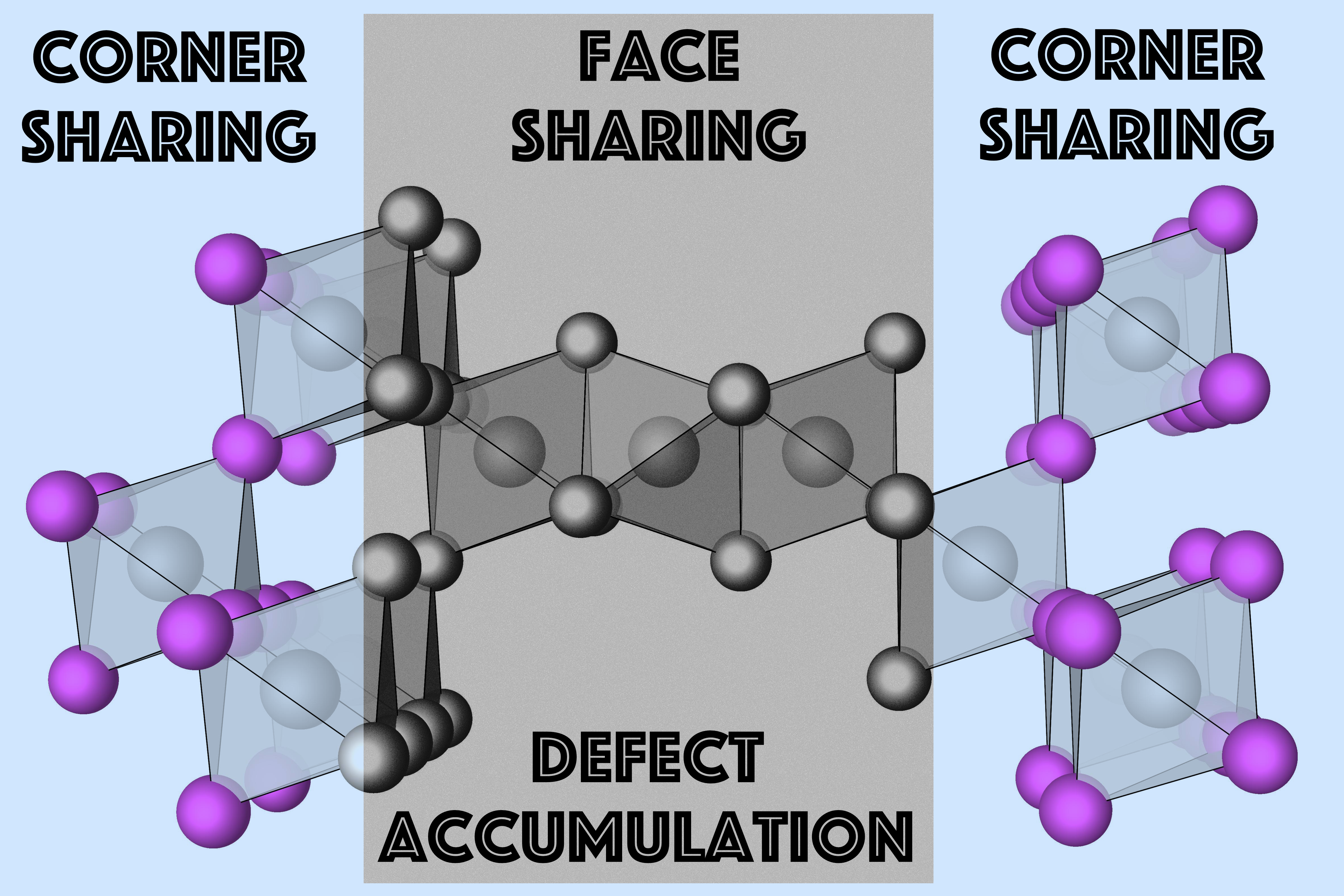} 
\end{tocentry}
%
% 5 cm x 7.5 cm
%

\begin{abstract}
The competition between corner, edge and face-sharing octahedral networks is a cause of phase inhomogeneity in metal halide perovskite thin-films. Here we probe the charged iodine vacancy distribution and transport at the junction between cubic and hexagonal polytypes of CsPbI$_3$ from first-principles materials modelling. We predict a lower defect formation energy in the face-sharing regions, which correlates with a longer Pb--I bond length and causes a million-fold increase in local defect concentration. These defects are predicted to be more mobile in the face-sharing regions with a reduced activation energy for vacancy-mediated diffusion. We conclude that hexagonal phase inclusions or interfaces will act as defect sinks that could trap charges and enhance current-voltage hysteresis in perovskite-based solar cells and electrical devices. 
\end{abstract}


%\newpage

%Suggested paper plan for ACS Energy Letters:
%\begin{enumerate}
%\item Summarise recent evidence of mixed phase perovskites including Sam Stranks (Science, 2021) 
%\item Figure 1 - polytype structures (AW can make to be consistent with our recent papers)
%\item Analyse defect formation energy trends and correlation to bond length
%\item Figure 2 - the four panels of defect formation and concentrations in the two polytypes
%\item Figure 3 - trends of other possible descriptors
%\item Analyse activation energy for defect transport in 3C vs 2H. Do we have the polytype numbers?
%\item Discussion - implications for defect concentrations - enhance recombination and transport. Tie to experimental observations.
%\item SI - octahedral distortion upon relaxation figure. dielectric constant table and analysis. 
%\end{enumerate}

%\newpage
%\item Summarise recent evidence of mixed phase perovskites including Sam Stranks (Science, 2021) 

%\section{Introduction}

 Metal halide perovskites are promising materials for photovoltaic and optoelectronic applications.\cite{kim2020high} Single-junction metal halide perovskite solar cells have achieved power conversion efficiencies of more than 25\% in just over a decade since their first report.\cite{kojima2009organometal,yang2015high,min2019efficient,url:NREL} However, structural transformations from photoactive corner-sharing phases (e.g. cubic or tetragonal) to photoinactive edge or face-sharing phases are a cause of degradation.\cite{stoumpos2013semiconducting,tan2020shallow,an2021structural} Beyond simple phase mixing, a range of ordered polytype structures can be formed.\cite{stoumpos2017structure,park2020hexagonal,tian2021progressive,li2021evolutionary}
 Interesting phenomena can emerge at the interface between regions of different connectivity.\cite{pavlovetc2020suppressing,fop2020high,li2021critical} For example, fast ion diffusion has been observed between corner and face-sharing regions of 6H metal oxide polytype derivatives, which is linked to a high concentration of defects.\cite{sinclair1999structure}
 
There is an increasing amount of evidence that hexagonal phase inclusions are ubiquitous in metal halide perovskite films, even those with high photovoltaic performance.\cite{marchezi2020degradation}
For example, the presence of 2H polytype domains have been associated with high trap concentrations from photoemission electron microscopy.\cite{macpherson2022local} 
The growing consensus is that domains of such secondary phases can generate clusters of sub-bandgap states that are detrimental to photovoltaic performance and induce photodegradation under operational conditions.\cite{kosar2021unraveling,nan2021revealing,jones2019lattice,doherty2020performance,tan2020shallow} However, the underlying origin and mechanism for such behavior has not been identified. 

\begin{figure}
  \centering
    \includegraphics[width=12cm]{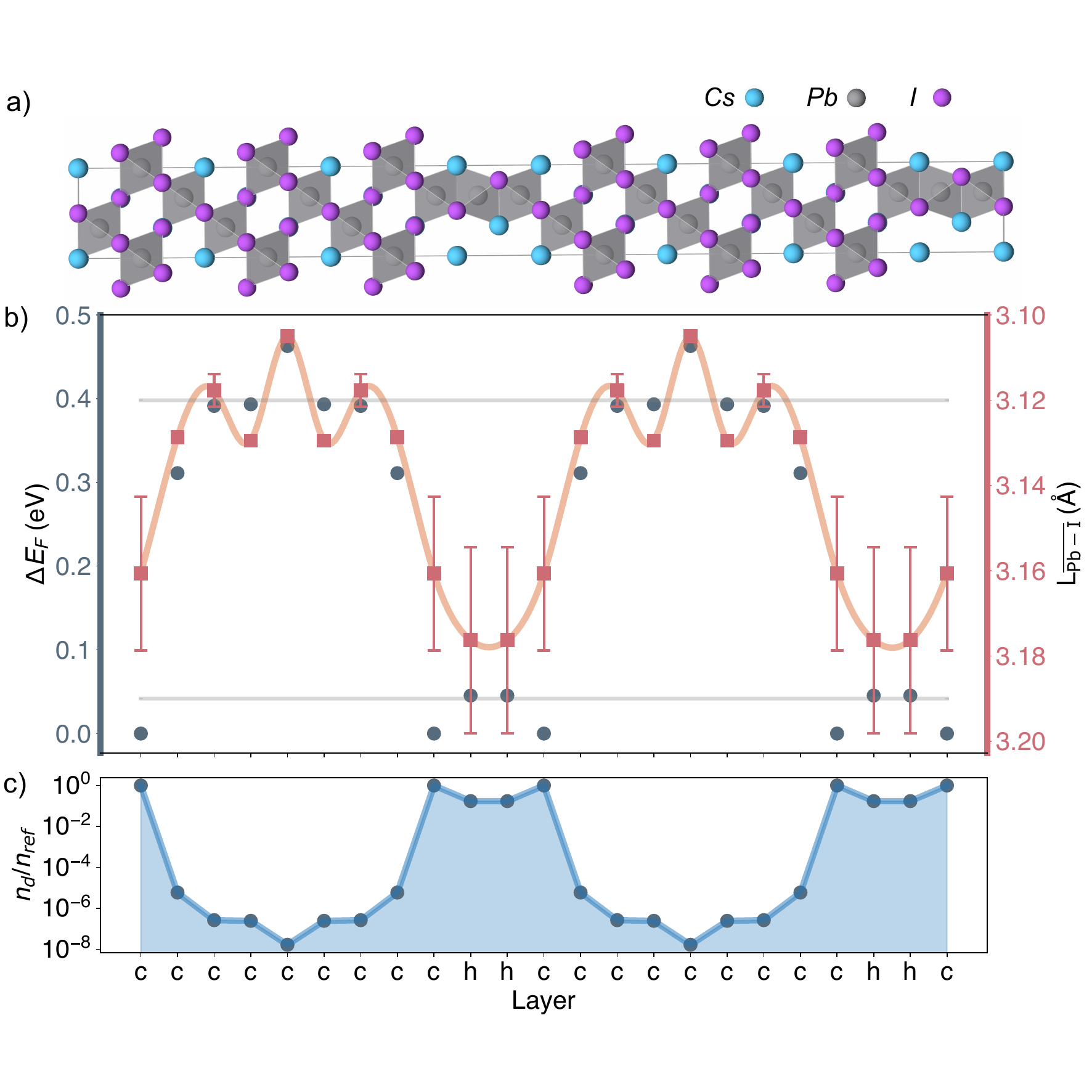}
    \caption{
    (a) Illustration of the 11H 2$h$9$c$ polytype structure of \ce{CsPbI3}. The lead iodide octahedra are shaded gray. Note that the stacking axis corresponds to the ⟨111⟩ direction for a cubic perovskite unit cell. (b) Relative defect formation energy ($\Delta E_{F}$, blue dots) and average Pb--I length ($\Delta\mathrm{L}_{\overline{\mathrm{Pb-I}}}$, red squares) along the stacking layers. The grey horizontal lines represent the average Pb--I bond length in the perfect cubic (3.12 Å) and hexagonal (3.19 Å) structures. The error bars denote the deviation from the average Pb--I length in each octahedron; the curved line is drawn to guide the eye. (c) The relative defect concentration assuming thermodynamic equilibrium at $T = 300$ K. 
}
  \label{f1}
\end{figure}

In this work, we report an investigation of the charged halide vacancy distribution and transport at the interface between corner and face-sharing regions of \ce{CsPbI3} using first-principles materials simulations. We model the defect processes using a representative 11H polytype that consists of two face-sharing layers (2$h$) connected to nine corner-sharing layers (9$c$) as shown in Figure \ref{f1}a. For comparison, results for the alternative 11H 3$c$8$h$ polytype are included as Supporting Information. We demonstrate that hexagonal phase inclusions or interfaces act as defect sinks in metal halide perovskites with mobile anion vacancy defects. 
 
\textbf{Defect accumulation at interfacial and face-sharing regions.}
The optimised bulk crystal structures for the 11H polytypes of \ce{CsPbI3} were taken from a previous study on the bulk crystal properties.\cite{li2021evolutionary} 
To describe the formation of point defects, a supercell expansion of 2$\times$2$\times$1 formed of 220 atoms was used.
We identified all symmetry inequivalent iodine vacancy sites ($V_{\mathrm{I}}^+$) to reduce the complexity of the problem to just six distinct defect types. 
The associated defect formation energies were then combined to reconstruct the full energy landscape plotted in Figure \ref{f1}.

The computed $V_{\mathrm{I}}^+$ formation energies in the $2h9c$ structure are shown in Figure \ref{f1}b. 
To avoid setting a chemical potential for iodine, which requires consideration of the specific processing conditions and chemical environment that may involve gas, solid or liquid exchanges with the perovskite crystal, we set the lowest energy configuration to zero and compared the relative defect formation energies.
A spread of 0.5 eV is found for forming the same defect type in different coordination environments. 
A clear two-region behaviour is observed in the formation energies with a higher-energy group in the corner-sharing regions and a lower-energy group in the face-sharing regions.
An underlying correlation is found between the $V_{\mathrm{I}}^+$ defect formation energy and the Pb--I bond length.
The corner-sharing regions have bond lengths similar to those in the parent 3C cubic perovskite ($\approx$ 3.12 Å), while the face-sharing regions have shorter bond lengths similar to the parent 2H phase ($\approx$ 3.19 Å). Iodine vacancy formation is preferred in the regions of the polytype with bond elongation. 

To understand the impact of the defect formation energies on the iodine vacancy defect distributions, the equilibrium concentrations were then calculated. 
The absolute concentration for a single defect species ($n_d$) is defined by the standard equilibrium expression
\begin{equation}
n_{d}=N_{\text {site}} \exp \left(-\frac{\Delta E_{F}}{k_{\mathrm{B}} T}\right)
\end{equation}
where $N_{\text{site}}$ represents the number of available crystal sites, $\Delta E_{F}$ is the defect formation energy, $k_B$ is the Boltzmann constant, and $T$ is temperature.\cite{kim2020quick}
We define the relative defect concentration ($n_{d}$/$n_{ref}$) as:
\begin{equation}
\frac{n_{d}}{n_{{ref}}}=\exp \left(-\frac{\Delta E_{F}-\Delta E_{ref}}{k_{B} T}\right)
\end{equation}
where $n_{ref}$ refers to the concentration at the site with lowest formation energy $\Delta E_{ref}$. 
The distributions are plotted in Figure \ref{f1}c, which highlight fluctuations over many orders of magnitude.
A high vacancy concentration is found around the face-sharing region, which is $\approx$ 10$^6$ times higher than the plateau in the corner-sharing region due to the exponential dependence on the underlying formation energy. 
The highest overall concentration is found in the interfacial layer, which is strained with bond length values that are shorter than the face-sharing regions and longer than the corner-sharing regions.
Based on these results, we conclude that hexagonal phase inclusions or interfaces are likely to act as sinks for vacancy defects in halide perovksites. 
The predictions are consistent with the recent report by Macpherson $et$ $al$\cite{macpherson2022local} that highlighted the high concentration of structural defects at the polytype junction, consistent with vacancies, which degrade device performance.

\textbf{Microscopic origin of the $V_{\mathrm{I}}^+$ defect distribution.}
Strain is an important parameter in the property control of metal halide perovskites. 
It is an intrinsic feature of perovskite-derived polytypes owing to the combination of building blocks with differing connectivity in the same crystal.\cite{jones2019lattice}
In particular, large changes in Pb--I bond lengths are found along the stacking axis of mixed polytypes.
The lowest defect formation energy is calculated at the stacking interface with expanded cubic and compressed hexagonal layers. 
For the case of homogeneous hydrostatic strain in a purely corner-sharing perovskite, lattice expansion makes iodine vacancy formation less favourable.\cite{jung2022pressure} 
The behaviour is different here as the strain is non-uniform with an elastic dipole centred around the interfacial iodide layer that reduces the defect formation energy. 

\begin{figure}
  \centering
    \includegraphics[width=10cm]{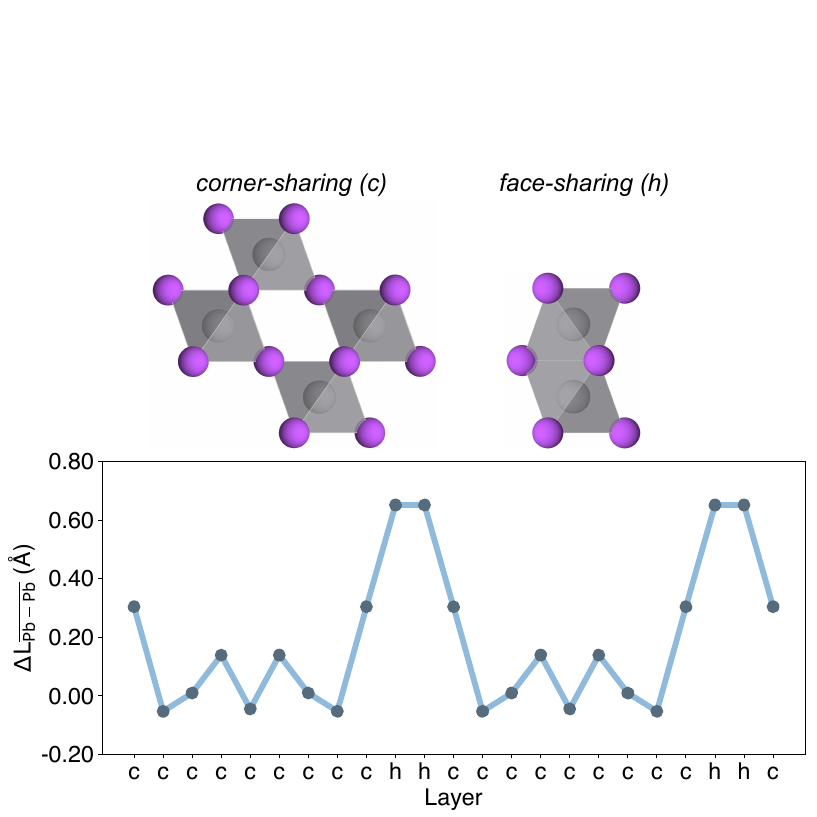}
    \caption{Pb--Pb bond length changes ($\Delta$$\mathrm{L}_{\overline{\mathrm{Pb-Pb}}}$, blue dots) following structural relaxation with a positively charged iodine defect and the pristine structure of the 11H ($2h9c$) polytype.
}
  \label{f2}
\end{figure}

Another factor that contributes to the vacancy formation energy is the underlying atomic rearrangements upon defect formation.
After removing an I$^-$ ion from the crystal, the electrostatic balance in the Pb$^{2+}$--I$^{-}$--Pb$^{2+}$ units along the stacking axis is replaced by a repulsive Pb$^{2+}$--$V_{\mathrm{I}}^+$--Pb$^{2+}$ interaction, which will force the pair of Pb$^{2+}$ cations in neighbouring octahedra away from each other, leading to elongated Pb--Pb distances. 
The correspond change in Pb separation is shown in Figure \ref{f2}. 
In both cubic and hexagonal regions, elongated Pb--Pb distances were found; however, the magnitude of the change in the hexagonal layers is much larger and exceeds 0.6 Å. 
In the interface region, an elongated Pb--Pb distance of 6.62 Å (from 6.31 Å) was observed with a 0.3 Å displacement of the Pb atoms. 
The correlation between the larger bond elongation in Figure \ref{f2} and lower defect formation energy in Figure \ref{f1}b indicates a stabilisation in the face-sharing regions that is likely driven by enhanced structural relaxation and the associated electrostatic stabilisation. 

\textbf{Charged vacancy transport.}
It is now well established that defects in metal halide perovskites are mobile, in particular on the anion sub-lattice.\cite{eames2015ionic,azpiroz2015defect,haruyama2015first,yang2016fast}
The magnitude of ionic transport depends on the concentration of diffusing species and their activation energy, in addition to external factors such as light exposure and applied electric fields.\cite{woo2022factors,walsh2018taking,kim2018large}

\begin{figure}
  \centering
    \includegraphics[width=10cm]{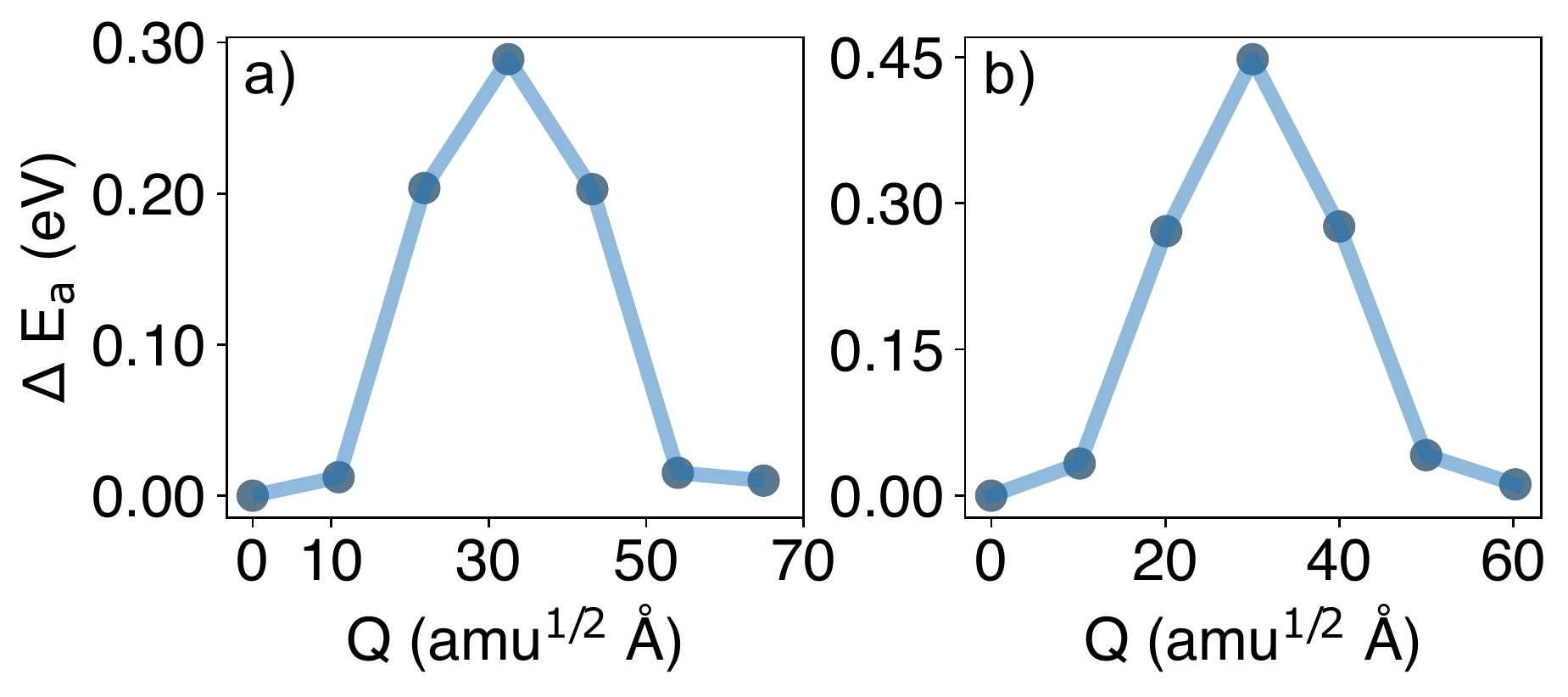}
    \caption{Calculated migration barrier ($\Delta$E$_a$) for $V_{\mathrm{I}}^+$ migration out-of-plane (a) and in-plane (b) in the hexagonal 2H phase of \ce{CsPbI3}. The corresponding barrier in the 3C phase is 0.34 eV at the same level of theory.
}
  \label{f3}
\end{figure}

We first compare the calculated ion migration barrier in the face-sharing (2H) and corner-sharing (3C)  structures.
A transition state search is performed using the nudged elastic band (NEB) method,\cite{mills1995reversible} with a climbing-image algorithm based on five images to estimate the diffusion barriers.
The barrier is isotropic in the 3C phase, but there is the possibility for in-plane and out-of-plane transport in the 2H phase (see Figure \ref{f3}).
The calculated activation energy of 0.34 eV (3C) decreases to 0.29 eV (2H) for migrating out-of-plane along the $\langle0001\rangle$ stacking axis.
This suggests that the face-sharing domains can support similarly high levels of ion transport. 
In contrast, the in-plane 2H barrier increases to 0.45 eV along $\langle1000\rangle$.

Mechanical stress and strain affect defect mobility and the energy barrier of halide ion migration increases with compression (applied pressure).\cite{woo2022factors,muscarella2020lattice} 
Therefore, at the interface with the 11H polytype, the ion migration barrier will be lowered due to the elongated Pb--I bond length (from 3.12 Å to 3.16 Å) in the expanded cubic region. 
The high defect concentration at the junction will also support a higher ion flux.
We noted that our calculated energy barriers for diffusion are in good agreement with experimental measurements, which is typically in the 0.2$-$0.5 eV range for vacancy-assisted halide ion diffusion in metal halide perovskites.\cite{iwahara2009ionic} 

In conclusion, we have demonstrated that hexagonal phase inclusions in metal halide perovskites will act as vacancy sinks due to a reduced formation energy in both the face-sharing regions and the interfaces between regions of different connectivity.
The behavior is linked to changes in the Pb--I bond lengths, which show a clear correlation to the energy cost to generate the vacancy defect along the two-region polytype. 
The associated equilibrium distribution of defects changes by orders of magnitude. 
Finally, we show the the face-sharing units have comparable activation energy for defect transport compared to the regular corner-sharing perovskites.

Our study provides several insights into observed behaviour of these materials and opens questions for further study. 
Firstly, we have established the preference for vacancy defects that are associated with current-voltage hysteresis; however, many other defect species can form such as isolated and aggregated interstitials.\cite{ambrosio2020formation}
It is clear that non-radiative pathways exist at the polytype junctions, \cite{kosar2021unraveling,nan2021revealing,doherty2020performance,tan2020shallow,macpherson2022local} but the exact combination of dilute or aggregated impurities remains to be identified. 
Also our modelling is focused on equilibrium distributions in the dark. Illumination may change the observed behaviour, including the generation of metastable defect configurations and non-equilibrium defect distributions driven by chemical potential changes.  
Deeper research on polytype interface stability, including the role of chemical composition, is required to complete our understanding. 

\section{Methods}
The crystals structures were generated using the \textsc{polytype} code available from \url{https://github.com/WMD-group/polytype}.
The pre-processing and symmetry analysis for the defect calculations was performed using \textsc{pymatgen} (\url{https://pymatgen.org}) and the \textsc{doped} code available from \url{https://github.com/SMTG-UCL/doped}. 

The first-principles total energy and forces were obtained from on Kohn-Sham density-functional theory\cite{kohn1965self,hohenberg1964inhomogeneous} as implemented in \textsc{VASP}.\cite{kresse1996efficient,kresse1996efficiency} 
The projector augmented-wave method\cite{kresse1999ultrasoft} was employed with the Perdew-Burke-Ernzerhof exchange-correlation functional revised for solids (PBEsol)\cite{perdew2008restoring} including scalar-relativistic effects.
For all calculations, the plane-wave kinetic energy cutoff was set to 500\,eV, while convergence criteria of $10^{-5}$\,eV and $10^{-2}$\,eV\AA$^{-1}$ for total energy and forces on each atoms, respectively, were employed.
For Brillouin zone sampling, $\varGamma$-centered \textit{k}-point meshes were set to 6$\times$6$\times$2 and 3$\times$3$\times$2 for geometry optimisation with primitive unit cells and supercells, respectively. 
For the charged defect calculations, anisotropic finite-sized corrections were included following Kumagai and Oba,\cite{kumagai2014electrostatics} as detailed in the Supporting Information. 
\begin{acknowledgement}
This research was supported by a grant of the Korea Health Technology R\&D Project through the Korea Health Industry Development Institute (KHIDI), funded by the Ministry of Health \& Welfare, Republic of Korea (grant number$\colon$HI19C1344). Computational resources have been provided by the KISTI Supercomputing Center (KSC-2021-CRE-0510). 
We are also grateful to the UK Materials and Molecular Modelling Hub for computational resources, which is partially funded by EPSRC (EP/P020194/1 and EP/T022213/1). 
\end{acknowledgement}

\begin{suppinfo}
The Supporting Information is available free of charge at https://pubs.acs.org/doi/xxxx.
Additional details on the defect correction scheme, dielectric constants, and analysis of the alternative 11H (3$c$8$h$) polymorph.

\textbf{
Data Availability Statement}

Data produced during this work is freely available in a repository at: \url{https://doi.org/10.5281/zenodo.xxxx}.
\end{suppinfo}

\section*{Author Contributions}
% https://casrai.org/credit/
The author contributions have been defined following the CRediT system.
Y.W.W.: Conceptualization, Investigation, Formal analysis, Methodology, Visualization, Writing – original draft. 
Z.L.: Methodology, Formal analysis, Writing – review \& editing. 
Y-K.J.: Methodology, Writing – review \& editing. 
J-S.P.: Methodology, Supervision, Writing – review \& editing. 
A.W.: Conceptualization, Methodology, Supervision, Writing – review \& editing.


\bibliography{lib}

\end{document}

\beginsupplement
%\section{Supplementary Information}
%\subsection{}

%
% 5 cm x 7.5 cm
%


\subsection{Dielectric response and charged defect corrections}

The formation of energy of the iodine vacancy defects is calculated from:
\begin{equation}
\Delta{E_{F}} = [E_{D,q} - E_{Bulk}] + q\mu_e +E_{corr}
\end{equation}
where E$_{D,q}$ and E$_{Bulk}$ are the energies of the charged defective and host supercells, respectively. $q$ and $\mu_e$ represent the charge state of the defect and the electronic chemical potential (Fermi level), respectively. 
E$_{corr}$ is a correction term for finite-size effects arising from the supercell approach. 
We employ the correction approach of Kumagai and Oba\cite{kumagai2014electrostatics} which account for the anisotropic dielectric screening of the long-ranged charged defect interactions. 

The electrostatic corrections required the calculated dielectric constants. 
These required a high density of \textit{k}-points with meshes of 12$\times$12$\times$12 used for the 2H and 3C. And 8$\times$8$\times$2 for the 11H polytypes.
The resulting values are shown in Table S1. 
The 11H polytypes with mixed connectivity
show large in-plane dielectric constants, more than double of the 2H polytype. 
The high built-in polarization, arising from the strained structure, is consistent with the report by Park et al.\cite{park2020hexagonal}

\begin{table*}[h]
\centering
\begin{tabular}{ccccccccccccccc}
 \hline\hline
 \noalign{\vskip 1mm}
 & \multicolumn{2}{c}{$\varepsilon_{xx}$}  & \multicolumn{2}{c}{$\varepsilon_{yy}$} & \multicolumn{2}{c}{$\varepsilon_{zz}$} \\ \cline{2-7}  
 
 & $\varepsilon_{opt}$ & $\varepsilon_{ion}$& $\varepsilon_{opt}$ & $\varepsilon_{ion}$ & $\varepsilon_{opt}$ & $\varepsilon_{ion}$  \\

 \noalign{\vskip 1mm}
 \hline
 \noalign{\vskip 1mm}
 2H &	\ce	4.4 & 17.4 & 4.4 & 17.4 & 4.4 & 44.8	\\ 
 \noalign{\vskip 1mm}
 11H (\ce{$2h9c$})&	\ce 4.6 & 42.8 & 4.6 & 42.8 & 4.8 & 26.0  \\
 \noalign{\vskip 1mm}
 11H (\ce{$3c8h$})&	\ce 4.7 & 37.3 & 4.7 & 37.3 & 4.7 & 25.4	\\
 \noalign{\vskip 1mm}
 3C &	\ce 5.3 & 28.7 & 5.3 & 28.7 & 5.3 & 28.7	\\
 \noalign{\vskip 1mm}
 \ce{MAPbI3}&	\ce 5.0\cite{glaser2015infrared} & 28.5\cite{sendner2016optical} &&&&	\\
 \ce{}&	\ce 5.5\cite{leguy2016experimental,valverde2015intrinsic}& 24.5\cite{anusca2017dielectric} &&&&	\\
 \ce{CsPbI3}&	\ce 5.3\cite{brgoch2014ab} & &&&&	\\
 \ce{CsPbBr3}&	\ce  & 20.5\cite{govinda2017behavior} &&&&	\\
 \noalign{\vskip 1mm}
 \hline\hline 
\end{tabular}
\caption{Calculated dielectric constants for \ce{CsPbI3} polytypes. 
Representative values from experimental and computational literature reports are also included.}
\label{tab:tbl1}
\end{table*}

\newpage

\subsection{Defect formation in the alternative 11H ($3c8h$) polytype}

We consider the alternative 11H $3c8h$ polytype for comparison to the $2h9c$ structure in the main text. 
The system itself is of less interest for solar energy conversion as it corresponds to an imperfect ``yellow phase'' with a larger band gap. 
All calculation settings were preserved.
The computed $V_{\mathrm{I}}^+$ formation energies in the $3c8h$ structure are shown in Figure S1.
Again the lowest defect formation energy is associated with the face-sharing regions of the polytype. 
However, the interfacial strain field is different and although the lowest defect formation energy is found in the centre of the hexagonal region, the energy increases towards the interface and produces a maximum in the penultimate face-sharing layer. 

\begin{figure}
  \centering
    \includegraphics[width=10cm]{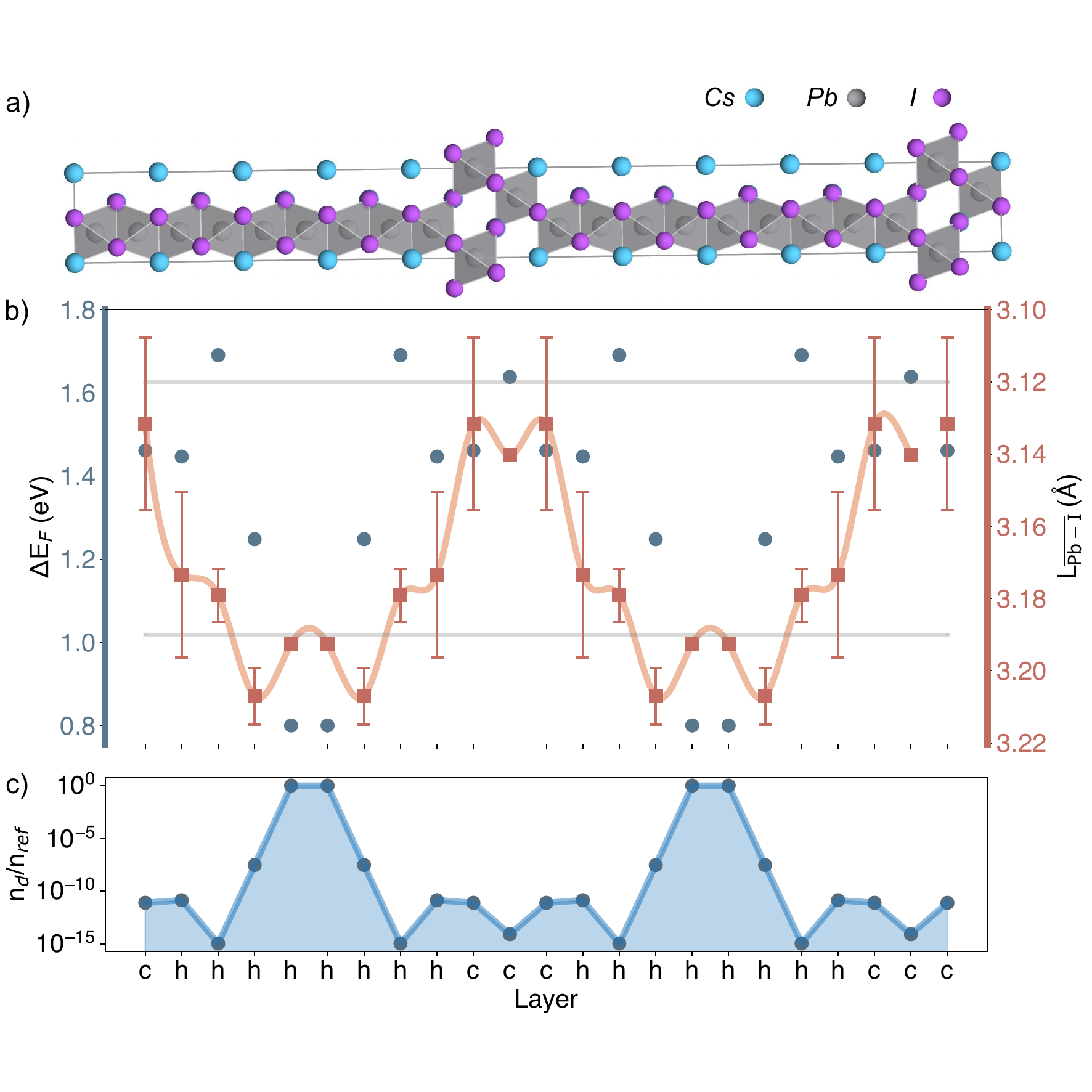}
    \caption{(a) Illustration of the 11H 3$c$8$h$ polytype structure of \ce{CsPbI3}. The lead iodide octahedra are shaded gray. Note that the stacking axis corresponds to the ⟨111⟩ direction for a cubic perovskite unit cell. (b) Relative defect formation energy ($\Delta$E$_F$, blue dots) and average Pb--I length ($\Delta\mathrm{L}_{\overline{\mathrm{Pb-I}}}$, red squares) along the stacking layers. The grey horizontal lines represent the average Pb--I bond length in the perfect cubic (3.12 Å) and hexagonal (3.19 Å) structures. The error bars denote the deviation from the average Pb--I length in each octahedron; the curved line is drawn to guide the eye. (c) The relative defect concentration assuming thermodynamic equilibrium at T = 300 K.
}
  \label{Figure1}
\end{figure}

\clearpage

\bibliography{./lib}